\documentstyle[epsf]{espart} 
\newcommand{\dspexp}[1]{\mbox{$e$\raisebox{2ex}{$\displaystyle{#1}$}}}

\def\approxge{\,\raisebox{-0.625ex}{$\stackrel{>}{\sim}$}\,}

\begin{document}
\mbox{}\hfill OCIP/C-97-01\\
\mbox{}\hfill nucl-th/9702032\\
\begin{frontmatter}
\title{Flux-Bubble Models and Mesonic Molecules}
\author{M.M.~Boyce, J.~Treurniet, and P.J.S. Watson}
\address{Ottawa-Carleton Institute for Physics, Carleton University,
Ottawa, Ontario, Canada, K1S-5B6}
\begin{abstract}
It has been shown  that the string-flip  potential model reproduces most
of the bulk properties of nuclear  matter, with the exception of nuclear
binding.   Furthermore,  it  was postulated  that   this  model with the
inclusion of the colour-hyperfine interaction should produce binding. In
some recent  work a modified version  of the string-flip potential model
was developed,  called the flux-bubble model,  which would allow for the
addition  of perturbative QCD interactions.  In  attempts to construct a
simple $q\bar q$ nucleon system using the  flux-bubble model (which only
included colour-Coulomb interactions)  difficulties arise with trying to
construct  a many-body  variational  wave function  that would take into
account  the locality of the  flux-bubble interactions. In this paper we
look at a toy system,  a mesonic molecule, in  order to understand these
difficulties.   {\it En route},   a   new variational  wave function  is
proposed   that may have   a sufficient  impact on  the  old string-flip
potential model  results that the inclusion  of perturbative effects may
not be needed.                                                          
\end{abstract}
\end{frontmatter}

\section{Introduction}

For the  past 30 years  several  attempts  have  been made, with  little
success, to describe nuclear matter in terms  of its constituent quarks.
The  main difficulty is due  to the non-perturbative  nature of QCD. The
only rigorous method for handling multi-quark systems to date is lattice
QCD.   However, this  is very  computationally intensive  and given  the
magnitude  of the problem  it appears unlikely to be  useful in the near
future.\footnote{Some very recent  advancements  have been made  in  the
area of lattice QCD that have reduced computation time by several orders
of magnitude.  ``Now what took hundreds of  Cray Supercomputer hours can
be done in only  a few hours on  a laptop  computer'' \cite{kn:Lepage}.}
As a result, more phenomenological means must be considered.

  A good  phenomenological  model should be  able  to at least reproduce
qualitatively   all the overall bulk  properties   of nuclear matter. In
particular,
\begin{itemize}
\item nucleon gas at low densities with no van der Waals forces
\item nucleon binding at higher densities
\item nucleon swelling and saturation of nuclear forces with increasing
density
\item quark gas at extremely high densities
\end{itemize}
There are several models that attempt  to reproduce these properties but
none of them  does so  completely. In this  paper, only  the string-flip
potential    model~\cite{kn:Boyce,kn:HorowitzI,kn:Watson},  will      be
considered.  This  model appears to be  promising  because it reproduces
most of the  aforementioned  properties  with the exception  of  nucleon
binding: instead of  producing a  binding  of  about  8 MeV  at  nuclear
density,   one  finds anti-binding  of  about  25  MeV. However, various
efffects have been ignored, such as colour Coulomb and hyperfine forces,
relativistic  effects and many-quark clusters. In  this paper we attempt
to   understand   why,    in   terms of     a   toy mesonic    molecular
system~\cite{kn:thesisb}.

\subsection{The String-Flip Potential Model}
\label{sc:foo}

\begin{figure}[ht]
  \begin{center}
   \mbox{\epsfxsize=10cm
      \epsffile{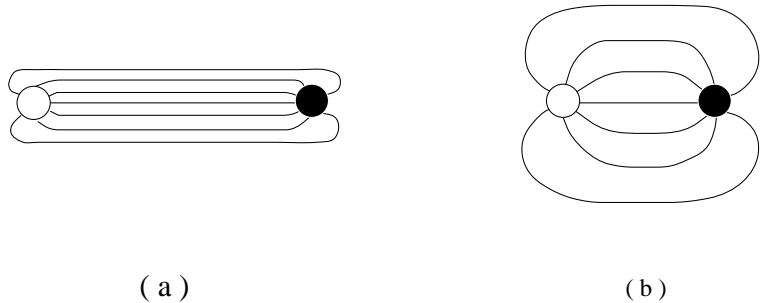}}
  \end{center}
\caption[$q\bar  q$  flux-tube diagram]{\footnotesize  The  colour field
     lines between quarks collapse  upon   themselves, due to the   self
     interacting    nature of the   gluons,    to form a  flux-tube-like
     structure.   At  long distances (a) the   fields  lines collapse to
     become  almost string-like and   at short distances  (b) the fields
     lines expand to become almost QED-like.}
\label{fig:qqpair}
\end{figure}

 The  idea of string-flip potential  models derives from lattice QCD and
meson spectroscopy.  A  static  potential derived from  computations  in
lattice QCD
\begin{equation}
V(r)\sim\sigma\,r-\frac{4}{3}\,\frac{\alpha_s}{r}\;,
\label{eq:phenpot}
\end{equation}
is   confirmed   as a  phenomenologically satisfactory   potential model
between quarks   and  antiquarks by   fitting  the  experimental mesonic
spectra.  This is   basically an interpolation  between  the  long range
non-perturbative      ($\sigma   r$)  and     short  range  perturbative
($-\frac{4}{3}\, \frac{\alpha_s}{r}$) parts  of the  force between pairs
of  quarks   (Fig.~\ref{fig:qqpair}).    In  the many-body   case,   the
string-flip   potential model ignores   the short   range  part of   the
potential  and   considers   an  ensemble   of  quark-antiquark   pairs,
$q\bar{q}$, such that the total string length, $\sum r_{q\bar{q}}\,$, is
a minimum: {\it i.e.},
\begin{equation}
V=\sigma\sum_{{\rm min}\{q\bar q\}}r_{q\bar q}\,.\label{eq:many}
\end{equation}

   This particular model has been used in an attempt to simulate nuclear
matter.   Although  it has an obvious   shortcoming, in that it  is more
applicable to  a pion gas, it  does surprisingly well at predicting some
of      the    overall           bulk   properties     of        nuclear
matter~\cite{kn:HorowitzI,kn:Watson}.   

\begin{figure}[ht]
\begin{center}
\mbox{\epsfysize=2cm\epsffile{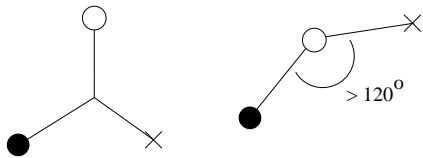}}
\end{center}
\caption[Flux-tube  arrangements for   the 3q cluster  potentials]{
   \footnotesize Flux-tube arrangements for the 3q cluster potentials.}
\label{fg:figa}
\end{figure}

   In a  previous paper, we  have extended this simple  model to  a more
realistic one  which involves triplets of  quarks.   Here the flux-tubes
leaving each quark meets  at a central  vertex such that overall length,
$r$, of ``flux-tubing''  is  minimized ({\it cf}.   Fig.~\ref{fg:figa}).
The  potential  energy is simply   $\sigma r\,$~\cite{kn:CarlsonB}.  The
effects of this are surprisingly small in going from $SU(2)$ to $SU(3)$:
a  nucleon  gas still forms, and  becomes  a quark   gas above a certain
critical density.

  These  models are motivated    by   results from lattice  QCD,   where
variations are taken about minimal  lattice field configurations between
quarks.   We  will refer  to these,   ``linear  potential  models,'' as,
``$SU_\ell(N)$  models,'' where the  $SU(N)$   refers to the   $SU_c(N)$
Yang-Mills gauge    group  and the  subscript,   ``$\ell$,'' refers  to,
``linear potential,''  \cite{kn:Boyce}.  In some   models each quark has
fixed   colour  to   simplify    the  searches  for  minimal   flux-tube
configurations as the  system  of quarks  evolves.  When  necessary, for
clarity, these  models  will  be referred  to  as, ``$SU_\ell^\prime(N)$
models.''

  This paper  is  divided into four sections:  the  first discusses  the
problems   faced when  attempts   are made   to extend the   string-flip
potential  model to include  perturbative QCD interactions for many-body
systems.  This is followed by a section  on mesonic molecules, where the
problems faced with the string-flip  potential model and its  extensions
are examined in detail. In the next we examine a modified wave function,
which  appears  to solve  many of   the difficulties.  The final section
reviews all of the findings and their possible consequences for modeling
nuclear matter.

\section{The Flux-Bubble Model}

  It appears  from earlier work that the  String-Flip potential model is
incapable  of  producing nuclear  binding   without the addition  of new
interactions ~\cite{kn:Boyce,kn:BoyceA}. However,   it turns out to   be
surprisingly difficult to  introduce these  into  a many-body  model.  A
model     was   recently      proposed,    called   the      Flux-Bubble
Model~\cite{kn:BoyceA}, which does allow the perturbative QCD effects to
be included.

  The primary objective is to construct a  model which combines both the
nonperturbative   (flux-tubes) and    perturbative (one-gluon  exchange)
aspects of QCD in  a consistent fashion. In order  to simplify this task
only colour-Coulomb extensions to a linear potential model using $SU(2)$
colour, {\it i.e.}, $SU_\ell(2)$,  will be considered.  The conventional
phenomenological potential,

\noindent
\mbox{}
\hfill ${\displaystyle
V(r)\sim\sigma\,r-\frac{4}{3}\,\frac{\alpha_s}{r}\;}$,
\hfill(\ref{eq:phenpot})
\vspace{1ex}

\noindent
mentioned in  \S~\ref{sc:foo} does   not extend naturally   to many-body
systems, since the $4/3$ must be replaced by $\lambda_{ij}=-3/4,1/4$ for
unlike and like colours respectively. This gives rise to a Van der Waals
potential between  colourless nucleons with a $\frac{1}{r^4}$ behaviour.
The flux bubble modifies this to
\begin{equation}
\begin{array}{c@{\;\;\;\;\;\;}c}
 {\rm v}_{ij}\sim\left\{
 \begin{array}{ll}
   {\displaystyle\sigma({\rm r}_{ij}-{\rm r}_0)} &
    {\displaystyle{\rm if}\;{\rm r}_{ij}>{\rm r}_0} \\
   {\displaystyle\alpha_s\lambda_{ij}
    \left(\frac{1}{\rm r_{ij}}-\frac{1}{{\rm r}_0}\right)} &
    {\displaystyle\mbox {\rm if}\;{\rm r}_{ij}<{\rm r}_0}
 \end{array}
 \right.
&
  \raisebox{-0.75cm}{\epsfxsize=2.5cm
   \epsffile{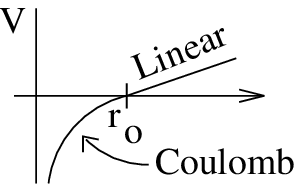}
  }
\end{array}
\end{equation}
with $\alpha_s\approx0.1\,$.   The    major difference    is that    the
nonperturbative and perturbative regimes are completely separated in the
latter, so that there are no Van Der Waals forces.\footnote{We have also
considered   a ``smoothed''  potential   where the  Coulomb term has  an
exponential cutoff of the form $e^{-\frac{r}{r_0}}$, and the linear term
is turned on by  a similar function. This  makes a negligible difference
to  our results.} When  the quarks  are separated at  a distance greater
than r$_0$ the potential is purely linear  and when they are inside this
radius it is purely Coulomb. In effect, for distances less than r$_0$, a
``bubble'' is formed in which the quarks are free  to move around, in an
asymptotically free fashion.  In both distance regimes the net colour of
the system  is neutral, and  phenomenologically  the models  are  almost
identical for a single $q\bar q$ system.

\begin{figure}[ht]
\begin{center}
\framebox{\setlength{\unitlength}{1mm}
\begin{picture}(130,62.5)
\put(-5,47.5){
   \setlength{\unitlength}{1mm}                   \begin{picture}(90,30)
   \put(0,0){\begin{minipage}{90mm}\caption[The   flux-bubble    model]{
   \footnotesize Consider  configuration (a) of   quarks, with r  $>{\rm
   r}_0$, about to move to  (b), s.t.  two  of them are within r  $<{\rm
   r}_0$.  Then  the procedure is to draw  a bubble of  ${\rm r}_0$ away
   from the two,  (b), and  to cut the  flux-tubes at  the boundary  and
   insert virtual $q\bar q$ pairs, (c).  Once  the potential is computed
   the configuration  is   restored to  (b)  before  the next   move  is
   made.}\label{fig:fluxbubble}\end{minipage}}\end{picture}}
\put(0,0){\mbox{\epsfxsize=130mm
   \epsffile{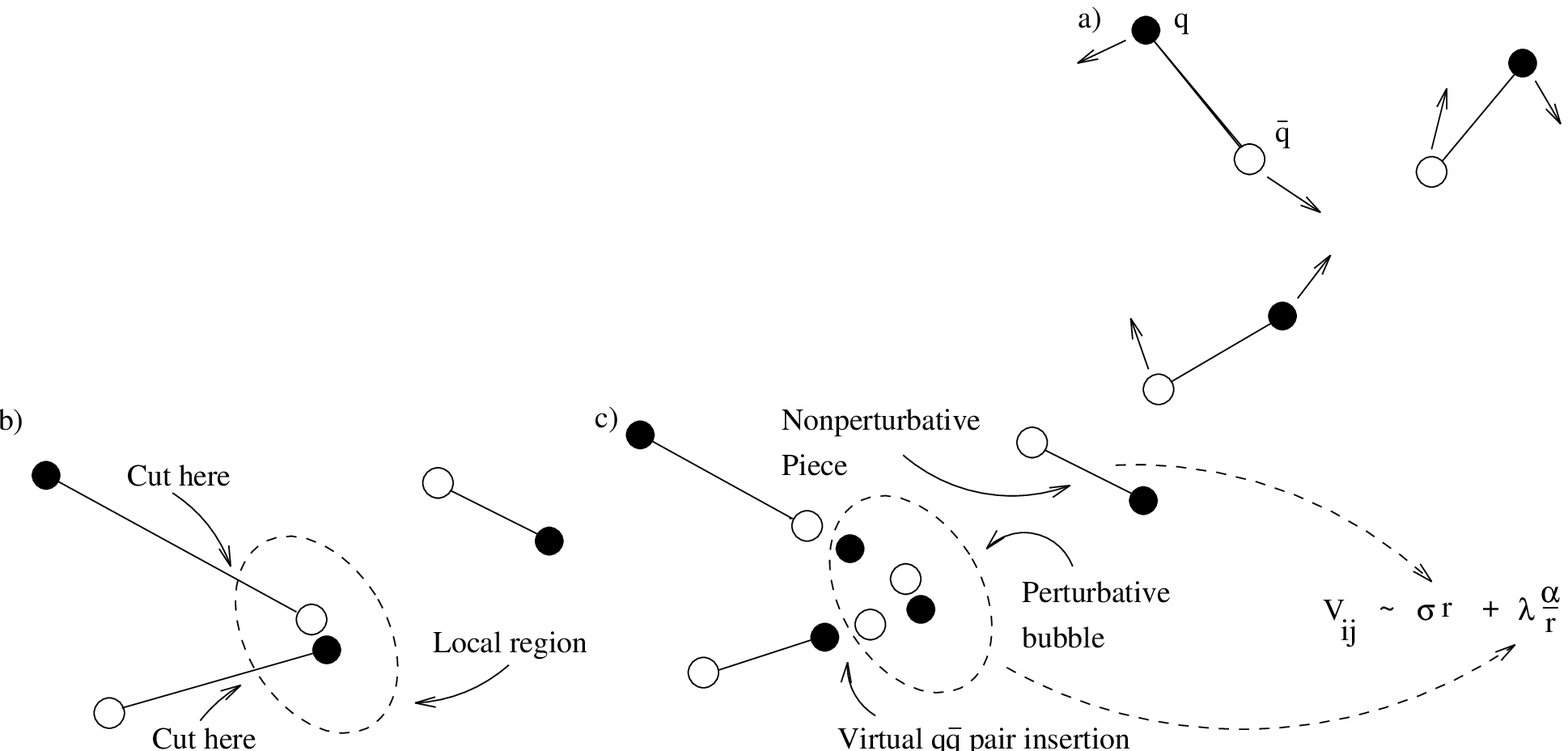}}}
\end{picture}}
\end{center}
\end{figure}

 This  extension of the linear potential,  although simple for a pair of
quarks, becomes more complex when considering  many pairs of quarks.  In
particular, it is difficult to construct a potential model when some set
of quarks  lie within  $r_0$, without forming  a colour  singlet so that
they must be  connected to more distant  quarks by  flux-tubes.  An {\it
ansatz}  that satisfies  these  requirements  is  obtained by  inserting
virtual $q\bar q$ pairs across any of the intersection boundaries formed
by the flux-tubes with the bubbles.  Now the segments of flux-tubes that
lie  outside the bubbles remain intact  while the segments inside simply
dissolve.   Fig.~\ref{fig:fluxbubble}  illustrates the dynamics  of this
model.

  Note the   insertion  of  the virtual $q\bar    q$   pairs allows  the
construction of  colourless objects. These are  solely used as a tool to
calculate the overall length of the flux-tube correctly, and not used in
computing the Coulomb term however, as the field energy is already taken
into account by the ``real'' quarks inside the bubbles. In general, once
the bubbles have been determined, the flux-tubes must be reconfigured in
order to minimize the linear part  of the potential.  Although the model
is currently for $SU_\ell(2)$ it  would be straightforward to extend  it
to a full $SU_\ell(3)$ model with all the one-gluon exchange phenomena.

  In  the previous  case  for  the  $SU_\ell^\prime(3)$ model, given  in
ref.~\cite{kn:Boyce},   we    adopted   a  wave   function    with   two
``independent''  parameters,   $\rho$  (density)  and   $\beta$ (inverse
correlation length); {\it i.e.},
\begin{equation}
\Psi_{\alpha\beta}=
\underbrace{\mathrm{e}^{-{\displaystyle \sum_{\min\{q\bar q\}}
   (\beta r_{q\bar{q}})^\alpha}
}\mbox{\rule[-2.25ex]{0em}{5ex}\hspace*{-2ex}}}_{\mbox{
\footnotesize Correlation $\Leftrightarrow$ $\beta$}}
\mbox{\hspace*{1ex}}
\underbrace{\prod_{\mbox{\tiny colour}}|\Phi(r_{p_k})|}_{\mbox{
\footnotesize Slater $\Leftrightarrow$ $\rho$}}\,,
\label{eq:watold}
\end{equation}
where $\alpha$   was   fixed.\footnote{Here, $\alpha\approx1.75$,    and
$\alpha=2$ gives the harmonic  oscillator case.}  We were  able to use  a
scaling  trick to reduce  the number of variation  degrees of freedom to
one, since  $\rho$ and   $\beta$  varied parametrically  with  a  single
parameter, $\theta\,$, as
\begin{equation}
(\beta,\rho^{1/3})\;\longrightarrow\;\zeta(\theta)\,
(\cos\theta,\sin\theta)\,
\end{equation}
where  $[\zeta(\theta)]\sim  fm^{-1}$.  However,   for  the  flux-bubble
potential, this scaling transformation  is broken.  This extra degree of
freedom greatly   increases  the computation  time,   since we must  now
perform a 2-dimensional minimization.

  We  have, however,  been   able  to show  that   the  results  of  the
computation    are   qualitatively    similar    to   previous   results
\cite{kn:BoyceA}: in other words, the flux  bubble does not give rise to
additional binding.  However, to  obtain satisfactory results requires a
coarse $10\times10$ mesh  of points, in  $\rho$  and $\beta$, and  hence
18hrs  of CPU time on  an 8 node farm.  Clearly  this procedure would be
ridiculously slow if  more  parameters were to be  added.  Thus a  rapid
method of checking different wave functions, minimization procedures and
various aspects of the  string-flip and flux-bubble potential models  is
desirable.

\section {Mesonic Molecules}

\begin{figure}[ht]
\begin{center}
\mbox{\epsfxsize=10cm
   \epsffile{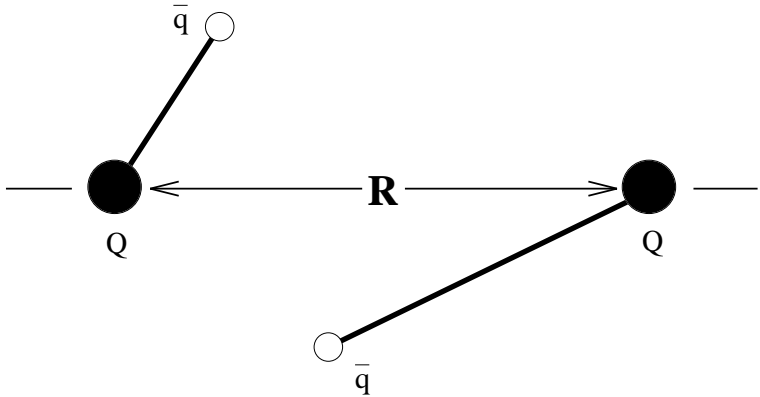}}
\end{center}
\caption[A mesonic-molecule]{
         A  mesonic-molecule,  $Q_2\,$, with  two  heavy quarks  and two
         light antiquarks.}
\label{fig:mesmol}
\end{figure}

   A   possible        mini-laboratory   is      a      mesonic-molecule
~\cite{kn:Weinstein},  $Q_2$, consisting of  two   heavy quarks and  two
relatively light  antiquarks: see Fig.~\ref{fig:mesmol}.The  quarks  are
assumed to be heavy so that the  light antiquarks can move around freely
without disturbing   their positions:  {\it  i.e.}, the Born-Oppenheimer
approximation is valid.   By varying the  distance, R, between the heavy
quarks a mesonic-molecular potential, $U(R)\,$, can be computed.
 
  The Schr\"odinger equation that describes  the effective potential,
$U(R)$, is given by \cite{kn:Schiff}
\begin{equation}
\left(\frac{1}{2m_q}
\sum_{\bar q}\vec\nabla^2_{\bar q}+V\right)\Psi=U(R)\Psi\,;
\label{eq:Schiff}
\end{equation}
we shall assume $m_q\approx330\,MeV$.   The potential $V$ describes  the
many-body  nature of  the  four  quark  system,  and is therefore  model
dependent.  This equation  can be solved  variationally for $\bar U(R)$,
at fixed values  of $R$, by guessing at  the form of the  wave function,
$\Psi$, and minimizing
\begin{equation}
\bar U=\bar T + \bar V\label{eq:effective}
\end{equation}
with respect to the parameters in $\Psi$.

 The  expectation values $\bar T$  and $\bar V$   are found by using the
Metropolis  algorithm  \cite{kn:Boyce,kn:Metropolis},  and   the optimal
parameters  for $\Psi$ are  found  by using the distributed minimization
algorithm   ~\cite{kn:thesisb,kn:BoyceC}  on  $\bar   U(R)$.   This  was
developed to handle the  problem of reducing  CPU overhead.  It involves
the usage of  several  $CPU$'s to perform interlaced  parabolic searches
for the minima  of a given surface.  For the $Q_2$ system this algorithm
yields
\begin{equation}
\tau_{\mbox{\tiny CPU}}^{\mbox{}}\;\sim\;{\cal O}(\frac{p}{m})\,,
\end{equation}
for $p$ parameters and $m$ computers: similar results are expected for
nuclear matter calculations. 

 The $Q_2$ system provides a  good way of  checking potential models for
the  possibility   of   nuclear  (mesonic) binding.      Because  of its
simplicity,  it  allows  for  investigating various   wave functions and
minimization schemes with minimal computational effort.

  A final useful feature is that the integration can be reduced to a 4-D
one if   Coulomb interactions are  ignored, and  a 5-D  one  if they are
included. This allows us to avoid the Metropolis procedure entirely, and
use multi-dimensional Gaussian   methods, which we  use a   check on the
Monte-Carlo methods

\subsection{Sensitivity to the Variational Parameters}

\label{sec-bkto}

  When the $SU_\ell(2)$ string-flip  potential model was investigated in
ref.~\cite{kn:Watson}  the parameter $\alpha$   in the  variational wave
function, given by Eq.~\ref{eq:watold},  was fixed by requiring  that it
minimize  the total energy at  zero density.   Since  this could be done
analytically it  allowed for  a  reduction in the number  of variational
parameters used in  the Monte Carlo.  It  was  assumed that constraining
$\alpha$ would have very little  effect on the  physics as a function of
density   since   the  results     only    vary  by   about    1\%   for
$1.5\,<\,\alpha\,<2.1\,$ at  zero density.  The  validity of  this claim
can now be checked more thoroughly by using the $Q_2$ mini-laboratory.

\begin{figure}[hbtp]
\begin{center}
   \mbox{\epsfxsize=135mm\epsffile{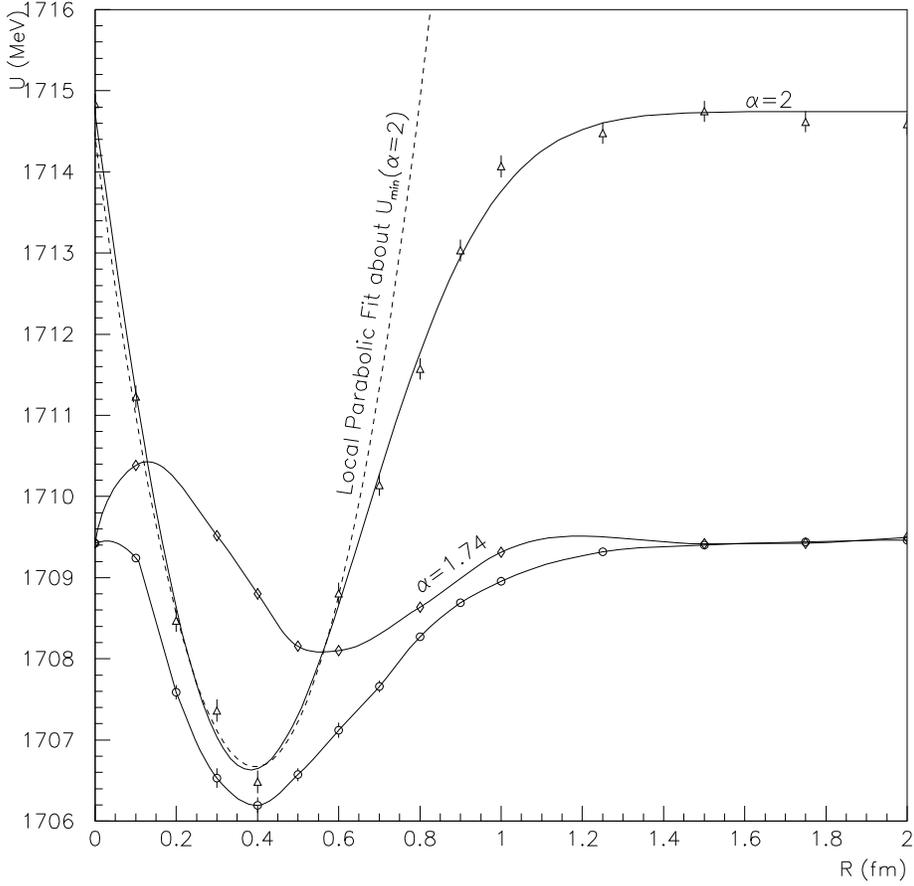}}
\end{center}
\caption[Graph of $SU_\ell(2)$ $Q_2$ potentials with various constraints
         on  $\alpha$]{\footnotesize  $\bar  U(R)$   where $\alpha$  is,
         allowed to vary, fixed at 2, and fixed at 1.74.}
\label{fig:wavetsta}
\end{figure}

  Fig.~\ref{fig:wavetsta} shows  a  plot of   the  Monte Carlo  results,
obtained  {\it  via}  Eqs.~\ref{eq:effective},   \ref{eq:watwav},    and
\ref{eq:linear}, for $\bar  U(R)$ where  $\alpha$  is allowed  to  vary,
$\alpha=2.00\,$, and $\alpha=1.74\,$.
The $\alpha=2.00\,$ value    was used in   the   old $SU_\ell(2)$  model
\cite{kn:Watson}, and  $\alpha\approx1.74$ was the  value that minimized
$\bar U(R)$ at infinite separation. The values of $\bar U(R)$ at the end
points of the curves, from $R=0\,fm$  and out to $R=5\,fm$, were checked
against the analytic solution given by Eq.~\ref{eq:anaa}.

  We find that when  $\alpha$ is allowed to  vary, $\alpha$  and $\beta$
become   highly  correlated.   Regardless,   looking   at   the  maximum
fluctuations about the  central   values  of the parameters    $\alpha$,
$\beta$,     and    $\bar      U(R)$    we     have    \cite{kn:thesisb}
$\alpha\approx(1.88\pm0.13)$, $\beta\approx(1.32\pm0.05)\,fm$, and $\bar
U(R)\approx(1710\pm4)\, MeV$, which is certainly  less then a 1\% effect
in  the total  energy   \cite{kn:Watson}. However  the relative   energy
$U(R)-U(\infty)$ depends critically on the variational procedure.

  Surprisingly  $\bar U(R)|_{\alpha=2}\,$ gives a  much deeper well than
if $\alpha$  was left as  a  free parameter: {\it i.e.},  $D\approx{\cal
O}(8)\,MeV$ \cite{kn:thesisb}.  For  certain fixed values of $\alpha\,$,
for example    $\alpha\approx1.74$   in  Fig.~\ref{fig:wavetsta},    the
potential  gives a slight  short   range repulsion, which becomes  quite
dramatic for $\alpha\approx1.0$. This is, of course, totally unphysical,
since $U(R)-U(\infty) \le 0$. We conclude from this that it is unsafe to
fix $\alpha\,$, since the well  depth can vary by nearly  a factor of 3.
However, it is fairly  safe to assume that the   results of past  papers
\cite{kn:Watson,kn:Boyce} will not  change significantly  if $\alpha$ is
allowed to vary.  For the rest  of the  paper, we consider  simultaneous
minimizations of $\alpha$ and $\beta\,$.

\subsection{A General Survey of Extensions to $SU_\ell(2)$}

\label{sec-old}

  In  this   section  the effects of    extending  the  old $SU_\ell(2)$
\cite{kn:Watson} model to  include flux-bubbles, with and  without fixed
colour  ({\it i.e.}, $SU_\ell(2)$ and $SU_\ell^\prime(2)$ respectively),
will be investigated in   the context of the  mesonic-molecular  system,
$Q_2\,$.  For  the old $SU_\ell(2)$  model the variational wave function
was assumed to be of the form \cite{kn:Watson},
\begin{equation}
\Psi_{\alpha\beta}=
\mathrm{e}^{
      -{\displaystyle \sum_{\min\{Q\bar q\}}
      (\beta r_{Q\bar{q}})^\alpha}
      }\,,\label{eq:watwav}
\end{equation}
where  $\alpha$   and    $\beta$   are  variational   parameters,    and
$r_{Q\bar{q}}$ is the distance   between  a given  $Q\bar q$   pair. The
summation in the exponent of the wave function is over the set of $Q\bar
q$ pairs that requires the least amount of flux-tubing: {\it i.e.},
\begin{equation}
V=\sigma\sum_{{\rm min}\{Q\bar q\}}r_{Q\bar q}
 \equiv\sigma\min\{\sum_{\{Q\bar q\}} r_{Q\bar q}\}\,,
 \label{eq:linear}
\end{equation}
or more explicitly
\begin{equation}
V=\sigma\min\{r_{Q_1\bar q_1}+r_{Q_2\bar q_2},\,
       r_{Q_1\bar q_2}+r_{Q_2\bar q_1}\}\,,
\end{equation}
{\it cf}. Eq.~(\ref{eq:many}). Therefore, the kinetic energy is simply
\begin{equation}
\bar T = \frac{\alpha\beta^\alpha}{2m_q}\,
 \langle
   \sum_{\min\{Q\bar q\}}[\alpha(1-(\beta r_{Q\bar q})^\alpha)+1]\,
   r_{Q\bar q}^{\alpha-2}
 \rangle\,.\label{eq:effectr}
\end{equation}

  Now,  if we assume  for the moment that  the colour is  affixed to the
quarks   ({\it  i.e.},  $SU_\ell^\prime(2)$),   then   the most  general
flux-bubble potential is of the form
\begin{equation}
V=\sigma\,\sum_{\min\{Q\bar q\}}
   (r_{Q\bar q}-r_0)\,\theta(r_{Q\bar q}-r_0)+
   \alpha_s
   \sum_{i<j}\lambda_{p_ip_j}
   \left(\frac{1}{r_{p_ip_j}}-\frac{1}{r_0}\right)
   \theta(r_0-r_{p_ip_j})
\,,\label{eq:stringyb}
\end{equation}
with  particle index $p_k\;\varepsilon\;\{Q_i,\bar q_j|i,j=1,2\}\,$,
such that $k=1,2,3,4\,$, and $SU_c(2)$ colour factor
\begin{equation}
\lambda_{p_ip_j}=\left\{\begin{array}{rl}
-\frac{3}{4}& {\rm if}\;p_ip_j\,
       \varepsilon\,\{\bar qQ\}\\
 \frac{1}{4}& {\rm if}\;p_ip_j\,
       \varepsilon\,\{\bar q\bar q,QQ\}
\end{array}\right.\,.
\end{equation}
This potential can be rewritten into the more enlightening form
\begin{eqnarray}
V&=&\sum_{\min\{q_i\bar q_j\}}
\left[
      \sigma\,(r_{q_i\bar q_j}-r_0)\,\theta(r_{q_i\bar q_j}-r_0)-
      \frac{3}{4}\,\alpha_s\,
      \left(\frac{1}{r_{q_i\bar q_j}}-\frac{1}{r_0}\right)
      \theta(r_0-r_{q_i\bar q_j})
\right]\nonumber\\&&
-\frac{3}{4}\,\alpha_s
\sum_{q_i\bar q_j\,\varepsilon\,\overline{\min\{q_i\bar q_j\}}_+}
\left(\frac{1}{r_{q_i\bar q_j}}-\frac{1}{r_0}\right)
\theta(r_0-r_{q_i\bar q_j})
\nonumber\\&&
+\frac{1}{4}\,\alpha_s
\sum_{q_iq_j\,\varepsilon\,\overline{\min\{q_i\bar q_j\}}_-}
\left(\frac{1}{r_{q_iq_j}}-\frac{1}{r_0}\right)
\theta(r_0-r_{q_iq_j})\,,\label{eq:fixed}
\end{eqnarray}
where  $q_i\,\varepsilon\,\{Q,q\}\,$.  The sets $\overline{\min\{q_i\bar
q_j\}}_+$ and $\overline{\min\{q_i\bar  q_j\}}_-$ contain the attractive
and  repulsive   quark    pairings, respectively,  in    the  complement
($\overline{\min\{q_i\bar q_j\}}$)  of the  set $\min\{q_i\bar q_j\}\,$:
{\it  i.e.}, $$ \overline{\min\{q_i\bar q_j\}}=  \overline{\min\{q_i\bar
q_j\}}_+ \cup \overline{\min\{q_i\bar   q_j\}}_-\,.\footnote{{\it e.g.},
if \rule[-2ex]{0em}{2ex} $\min\{q_i\bar   q_j\}=\{Q_1\bar  q_2,\,Q_2\bar
q_1\}$    then  $\overline{\min\{q_i\bar q_j\}}=\{Q_1\bar  q_1,\,Q_2\bar
q_2,\,Q_1      Q_2,\,\bar         q_1\bar    q_2\}$,   which     implies
$\overline{\min\{q_i\bar  q_j\}}_+=\{Q_1\bar  q_1,\,Q_2\bar q_2\}$   and
$\overline{\min\{q_i\bar  q_j\}}_-=\{Q_1 Q_2,\,\bar q_1\bar q_2\}$.}  $$
The    two terms inside    the  square  brackets  of  Eq.~\ref{eq:fixed}
represents a linear potential  followed by its colour-Coulomb extension:
{\it cf}.   Eq.~(\ref{eq:phenpot}).  The  remaining terms  represent the
rest of the colour-Coulomb  interactions, which contain both  attractive
and repulsive bits.

\begin{figure}[htbp]
\begin{center}
   \mbox{\epsfxsize=135mm\epsffile{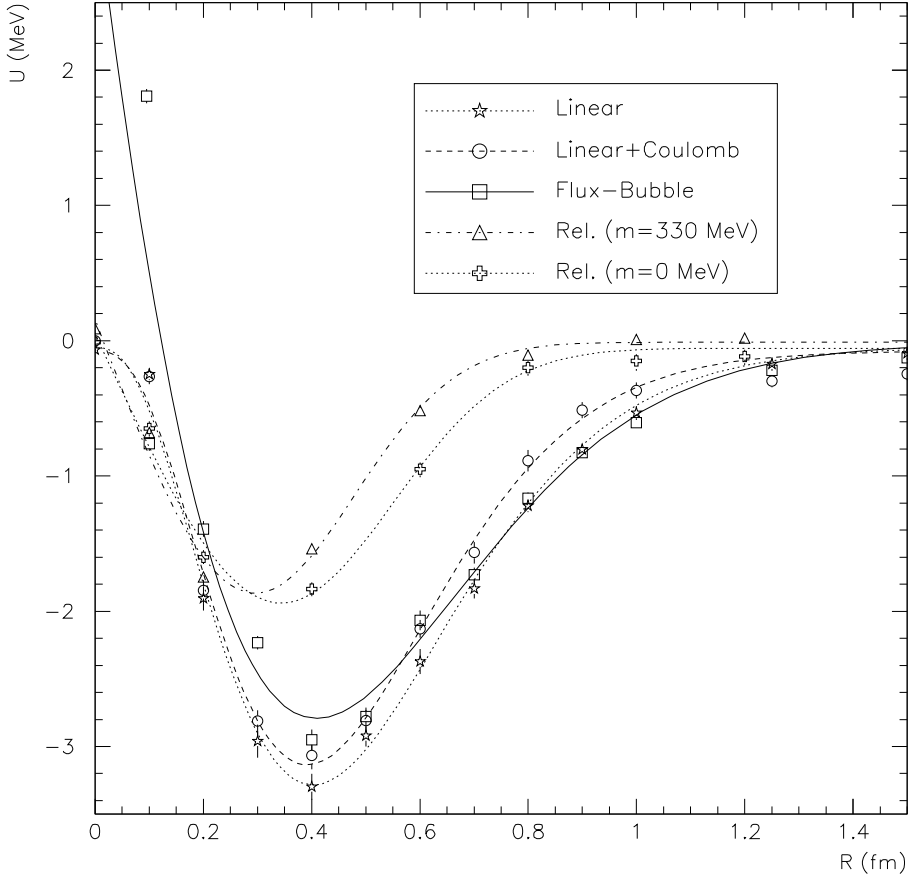}}
\end{center}
\caption[foo]{\footnotesize  The   binding        energy,       $E_B(R)$
    ($=U(R)-U_\infty$ s.t.  $U_\infty\equiv  U(\infty)$), for  the $Q_2$
    system    as function  of  heavy   quark   separation for  a  linear
    (Eq.~(\ref{eq:linear})), a   linear-plus-Coulomb (square brackets of
    Eq.~(\ref{eq:fixed})),  and  a  flux-bubble   (Eq.~(\ref{eq:fixed}))
    potential  model.    Also   included,    for  contrast,    are   the
    semi-relativistic   results   discussed in  \S~\ref{sec-chpdsc}. The
    curves in    these   plots have   been  parameterized  by  $u(r)\sim
    u_0\{e^{-2\kappa b(r-r_0)^a}   -       2\kappa   e^{-b(r-r_0)^a}\}(r
    -\varepsilon)^\eta$     ({\it          cf}.                    Morse
    potential~\cite{kn:Schiff,kn:Flugge}).}
\label{fig:oldwf}
\end{figure}

  Fig.~\ref{fig:oldwf}  shows the results   of  the Monte Carlo for  the
linear  potential and  its  various variants,  including the flux-bubble
potential.

 We   have also  considered  relativistic  effects,  via the relatvistic
Dirac-Hartree-Fock method({\it e.g.},~\cite{kn:Liberman}),  assuming the
flux-tube interaction is a Lorentz scalar.
\begin{equation}
\sum_{\min\{Q\bar q\}}[\nabla^2_{\bar q}+(m_q+\sigma r_{Q\bar q})^2]\Psi
=U^2(R)\Psi\,.
\label{fig:oneoverrsq}
\end{equation}
This   slightly exacerbates the problem, in   that the potential becomes
even more more shallow.  We note in passing that  it is not possible  to
solve the physically interesting case of massless quarks interacting via
a Coulomb potential in addition to  the linear term, since it introduces
an effective $1/r^2$ potential in Eq.~(\ref{fig:oneoverrsq}).

  To compare  these potentials with  more conventional methods, we would
like to  investigate whether they can  bind heavy  quarks into molecular
systems, and to   what extent they  resemble  conventional inter-nucleon
potentials.     The    details  of   this  analysis   can  be  found  in
appendix~\ref{sec-anal}.  A reduced heavy quark mass of
\begin{equation}
\mu_Q\approxge(53.7\pm1.9)\,GeV
\end{equation}
is  required to obtain binding for  these potentials. This should not be
too surprising as the potential wells are very shallow,
\begin{equation} 
\bar D\approx(2.986\pm0.030)\,MeV\,. 
\end{equation}

 To  find whether the  string-flip  potential models actually mimic pion
exchange,  the  asymptotic parts of the   $\bar U(R)$'s are  fitted to a
Yukawa  potential.  The results of the   analysis  yielded an effective
exchange mass of
\begin{equation}
\bar m_{\rm ex}\approx(636\pm41)\,MeV\,,
\label{eq:stockex}
\end{equation}
which is about 4.5 times too big, so the potential is much too
short-range.

\begin{figure}[htbp]
\begin{center}
   \mbox{\epsfxsize=135mm\epsffile{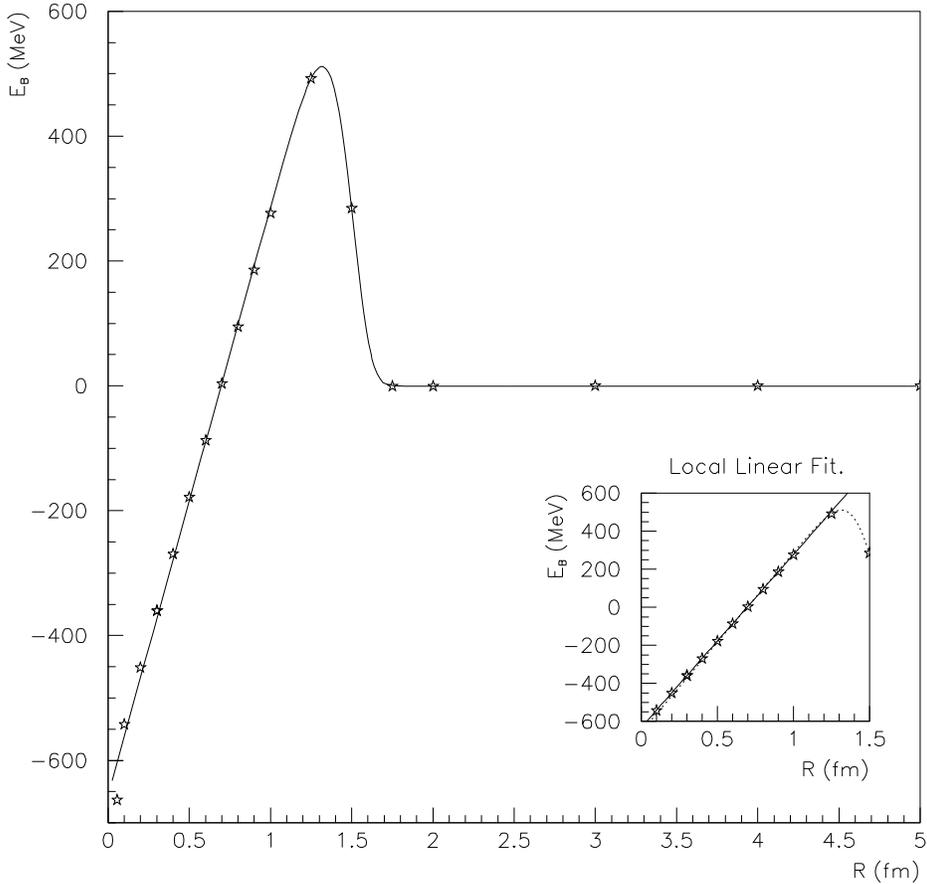}}
\end{center}
\caption[foo]{\footnotesize The   $Q_2$ binding energy curve   using the
         full flux-bubble interaction, {\it  i.e.}, in which the  colour
         is allowed to move around.   This curve was parameterized using
         $u(r)$ given in Fig.~\ref{fig:oldwf}.}
\label{fig:modelda}
\end{figure}

 Finally,  the flux-bubble potential is extended  to allow the colour to
move   around.  The potential is similar   to that of Eq.~\ref{eq:fixed}
except now the particle indices, $p_k$, carry colour degrees of freedom:
\begin{eqnarray}
   q_k&\varepsilon& \{R_i,\bar B_i,r_i,\bar b_i|i=1,2\}
    \;\subseteq\;SU_c(2)\,,\nonumber\\
\bar q_k&\varepsilon& \{\bar R_i,B_i,\bar r_i,b_i|i=1,2\}
    \;\subseteq\;\overline{SU_c(2)}\,,\nonumber
\end{eqnarray}
{\it i.e.},
$$
b\;\sim\;\bar   r\,,\,\mbox{\ldots,   etc.}\;\;\Leftrightarrow\;\;
SU_c(2)\;\sim\;\overline{SU_c(2)}\,,
$$
where the  capital case  letters represent the  heavy  quarks, the lower
case   letters  represent the light   quarks,  the  letters  $r$ and $R$
represent the red quarks, and the letters $b$ and $B$ represent the blue
quarks. Therefore,   the  $\min\{Q\bar  q\}$  in  Eq.~\ref{eq:fixed}  is
determined by Eq.~\ref{eq:linear} such that colour is no longer fixed to
a given quark.   Fig.~\ref{fig:modelda} shows the  results of  the Monte
Carlo for this potential.


  Obviously the  result of  this is  totally   different from  the fixed
colour   model.   For  $R\le1\,fm$,  the  potential  just reproduces the
potential between the heavy quarks: a linear fit to this region (see the
insert    in  Fig.~\ref{fig:modelda})     yields     a      slope     of
$(909.8\pm1.0)\,MeV/fm$!  Near   the origin  the  potential   behaves as
$-\frac{1}{r}$, which is due to  the Coulomb attraction between the  two
heavy quarks.  Beyond $R=1\,fm$ there  is a barrier  and beyond  this no
more  structure. This apparently   bizarre result  can be understood  as
follows:  if  two mesons,  each containing   a heavy  quark  and a light
anti-quark,  are  brought together from infinity   they initially feel a
repulsive force. However, it  then becomes energetically  preferable for
the two mesons to dissociate into one  meson containing two heavy quarks
and  another containing two  light anti-quarks.   This arises since  the
model  has   a   Pauli-Gursey    symmetry~\cite{kn:Gursey}:   quarks and
antiquarks have the same representations. Although potentially realistic
as a description of  meson-anti-meson interaction, the $SU_c(2)\,$  with
moving colour does not make a satisfactory model for nuclear matter.

  From  this  discussion,  it  is  clear  that  the extensions   to  the
string-flip potential model  to include the colour-Coulomb  interactions
have  essentially no effect on  the $Q_2$  potential, and certainly will
not give rise to nuclear binding. Furthermore, the string-flip potential
does  not give a long-range interaction  similar to  pion exchange. When
the  colour was allowed  to  move around  a highly  unphysical situation
occurs  which suggested  that there was   perhaps  a problem with  using
$SU_c(2)$ or with the variational wave function  itself --- perhaps even
both. In the next section we investigate an alternative wave-function.

\section{A New Wave Function}

\label{sec-new}

 The two  previous sections have shown that  modifying  the potential or
modifying the variational  procedure produces a relatively  small effect
on   the  $Q_2$ interaction.   It  therefore  seems  plausible  that the
fundamental problem lies in our  choice of the  wave-function. A hint is
to  note the similarity  between the $Q_2$  mesonic-molecular system and
$H_2$ molecular system. The key reason for  the molecular binding is the
screening  effect caused by  the electrons  which are for  the most part
localized in between the protons. This localization is achieved by using
a   variational wave  function  that  is a superposition   of the direct
product     of         two     ground     state       hydrogen    atoms,
\cite{kn:Schiff,kn:Bransden,kn:Heitler}
\begin{equation}
\Psi\sim\dspexp{-\beta(r_{e_1P_1}+r_{e_2P_2})}+
\dspexp{-\beta(r_{e_2P_1}+r_{e_1P_2})}\,.
\end{equation}
The  effect is that  of a ``bond-centred''  wave  function. Although the
$Q_2$ system is far removed from its $H_2$ cousin from a dynamical point
of view   and  the  motivations  for  achieving  localization  are quite
different, it would seem plausible to use a similar $ansatz$:
\begin{equation}
\Psi_{\alpha,\beta}=
\dspexp{-\beta^\alpha(r_{\bar q_1Q_1}^\alpha+r_{\bar q_2Q_2}^\alpha)}+
\dspexp{-\beta^\alpha(r_{\bar q_2Q_1}^\alpha+r_{\bar q_1Q_2}^\alpha)}\,.
\label{eq:psdo}
\end{equation}
If $\bar q_1Q_1$ and $\bar   q_2Q_2$ represent two separate mesons  then
the first term  represents  the internal  meson interactions  while  the
second term represents the  external meson interactions. Notice that the
external interactions shut off as the separation, $R\,$, between the two
heavy quarks becomes large,
\begin{equation}
\lim_{R\rightarrow\infty}\bar \Psi(R)=
\dspexp{-\beta^\alpha(r_{\bar q_1Q_1}^\alpha+r_{\bar q_2Q_2}^\alpha)}
\label{eq:new}
\end{equation}
which is  the desired property. Furthermore,  when  the light quarks are
close in space, $\Psi(R)$ is considerably enhanced.

  Using Eq.~\ref{eq:Schiff}  and  Eq.~\ref{eq:new}  the kinetic   energy
contribution is more complex  for  this wave-function, and  now  becomes
\cite{kn:Boyce,kn:thesisb,kn:Ceperley}
\begin{equation}
\bar T = 2\,\bar T_{-s}-\bar F^2\,,
\end{equation}
where
\begin{eqnarray}
\bar T_{-s}&=&\frac{-1}{4m_q}\sum_{\bar q}
       \langle\nabla_{\bar q}^2\ln\Psi\rangle\nonumber\\
&&\nonumber\\
&=&\frac{\alpha\beta^\alpha}{4m_q}\left\langle
\frac{(\alpha+1)}{\Psi}\left[
  (r_{\bar q_1Q_1}^{\alpha-2}+r_{\bar q_2Q_2}^{\alpha-2})\,
  \dspexp{-\beta^\alpha(r_{\bar q_1Q_1}^\alpha+r_{\bar q_2Q_2}^\alpha)}
  \right.\right.\nonumber\\\nonumber\\&&\left.+\,
  (r_{\bar q_2Q_1}^{\alpha-2}+r_{\bar q_1Q_2}^{\alpha-2})\,
  \dspexp{-\beta^\alpha(r_{\bar q_2Q_1}^\alpha+r_{\bar q_1Q_2}^\alpha)}
\right]\nonumber\\\nonumber\\&&+\,
\frac{\alpha\beta^\alpha}{\Psi^2}\,[
  (r_{\bar q_1Q_1}^2+r_{\bar q_1Q_2}^2-R^2)\,
  r_{\bar q_1Q_1}^{\alpha-2}\,r_{\bar q_1Q_2}^{\alpha-2}
  \nonumber\\\nonumber\\&&
  +(r_{\bar q_2Q_1}^2+r_{\bar q_2Q_2}^2-R^2)\,
  r_{\bar q_2Q_1}^{\alpha-2}\,r_{\bar q_2Q_2}^{\alpha-2}
  \nonumber\\\nonumber\\&&-\,
  (r_{\bar q_1Q_1}^{2\alpha-2}+r_{\bar q_2Q_2}^{2\alpha-2}+
  r_{\bar q_2Q_1}^{2\alpha-2}+r_{\bar q_1Q_2}^{2\alpha-2})
]\,\nonumber\\\nonumber\\&&\times\left.
  \dspexp{-\beta^\alpha(r_{\bar q_1Q_1}^\alpha+r_{\bar q_2Q_1}^\alpha
      +r_{\bar q_1Q_2}^\alpha+r_{\bar q_2Q_2}^\alpha)}
\right\rangle\,,
\end{eqnarray}
and
\begin{eqnarray}
\bar F^2&=&\frac{1}{2m_q}\sum_{\bar q}
      \langle(\nabla_{\bar q}\ln\Psi)^2\rangle\nonumber\\
&&\nonumber\\
&=&\frac{\alpha^2\beta^{2\alpha}}{2m_q}
\left\langle\frac{1}{\Psi^2}\,\left[
  (r_{\bar q_1Q_1}^{2\alpha-2}+r_{\bar q_2Q_2}^{2\alpha-2})\,
  \dspexp{-2\beta^\alpha(r_{\bar q_1Q_1}^\alpha+r_{\bar q_2Q_2}^\alpha)}
  \right.\right.\nonumber\\\nonumber\\&&\left.+\,
  (r_{\bar q_2Q_1}^{2\alpha-2}+r_{\bar q_1Q_2}^{2\alpha-2})\,
  \dspexp{-2\beta^\alpha(r_{\bar q_2Q_1}^\alpha+r_{\bar q_1Q_2}^\alpha)}
\right]\nonumber\\\nonumber\\&&+\,[
  (r_{\bar q_1Q_1}^2+r_{\bar q_1Q_2}^2-R^2)\,
  r_{\bar q_1Q_1}^{\alpha-2}\,r_{\bar q_1Q_2}^{\alpha-2}
  \nonumber\\\nonumber\\&&+
  (r_{\bar q_2Q_1}^2+r_{\bar q_2Q_2}^2-R^2)\,
  r_{\bar q_2Q_1}^{\alpha-2}\,r_{\bar q_2Q_2}^{\alpha-2}]
  \nonumber\\\nonumber\\&&\left.\times
  \dspexp{-\beta^\alpha(r_{\bar q_1Q_1}^\alpha+r_{\bar q_2Q_1}^\alpha
      +r_{\bar q_1Q_2}^\alpha+r_{\bar q_2Q_2}^\alpha)}
\right\rangle\,.
\end{eqnarray}
It serves as a good check of the computation that in the large $R$ limit
the  kinetic  energy reduces   to the  kinetic  energy for  the old wave
function in the same limit:
\begin{eqnarray}
\lim_{R\rightarrow\infty}\bar T(R)&=&
\frac{\alpha\beta^{\alpha}}{2m_q}\left\langle
(\alpha+1)\,
(r_{\bar q_1Q_1}^{\alpha-2}+r_{\bar q_2Q_2}^{\alpha-2})-
\alpha\beta^\alpha\,
(r_{\bar q_1Q_1}^{2\alpha-2}+r_{\bar q_2Q_2}^{2\alpha-2})
\right\rangle\nonumber\\&&\nonumber\\
&=&\frac{\alpha\beta^{\alpha}}{2m_q}\left\langle\sum_{i=1}^2[
\alpha(1-(\beta r_{\bar q_iQ_i})^\alpha)+1]\,r_{\bar q_iQ_i}^{\alpha-2}
]\right\rangle\,,
\end{eqnarray}
{\it cf}. Eq.~\ref{eq:effectr}. Also direct evaluation  of the RHS leads
to (assuming $\Psi$ is properly normalized)
\begin{equation}
\lim_{R\rightarrow\infty}\bar T(R)=\frac{g_T(\alpha)}{m_q}\,\beta^2\,,
\end{equation}
which is just the kinetic term for the analytic solution given by
Eq.~\ref{eq:anab}.

\begin{figure}[htbp]
\begin{center}
   \mbox{\epsfxsize=135mm\epsffile{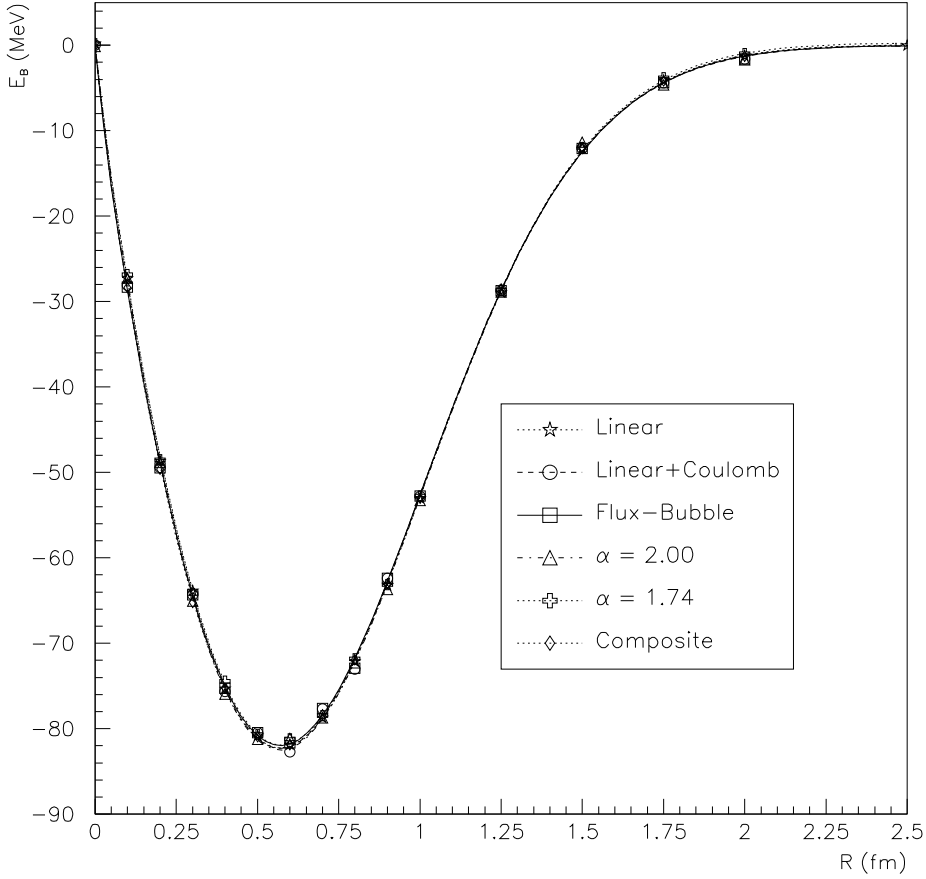}}
\end{center}
\caption[foo]{\footnotesize The binding energy,  $E_B(R)$, for the $Q_2$
    system  as function of   heavy  quark separation  for  a   linear, a
    linear-plus-Coulomb, and a flux-bubble potential model.  Also shown,
    are  the curves for   $\alpha=2.00$ and $1.74\,$, and the  composite
    model discussed  in \S~\ref{sec-chpdsc}. The  curves  in these plots
    have been parameterized by $u(r)$ given in Fig~\ref{fig:oldwf}.}
\label{fig:newwf}
\end{figure}

\begin{figure}[htbp]
\begin{center}
   \mbox{\epsfxsize=135mm\epsffile{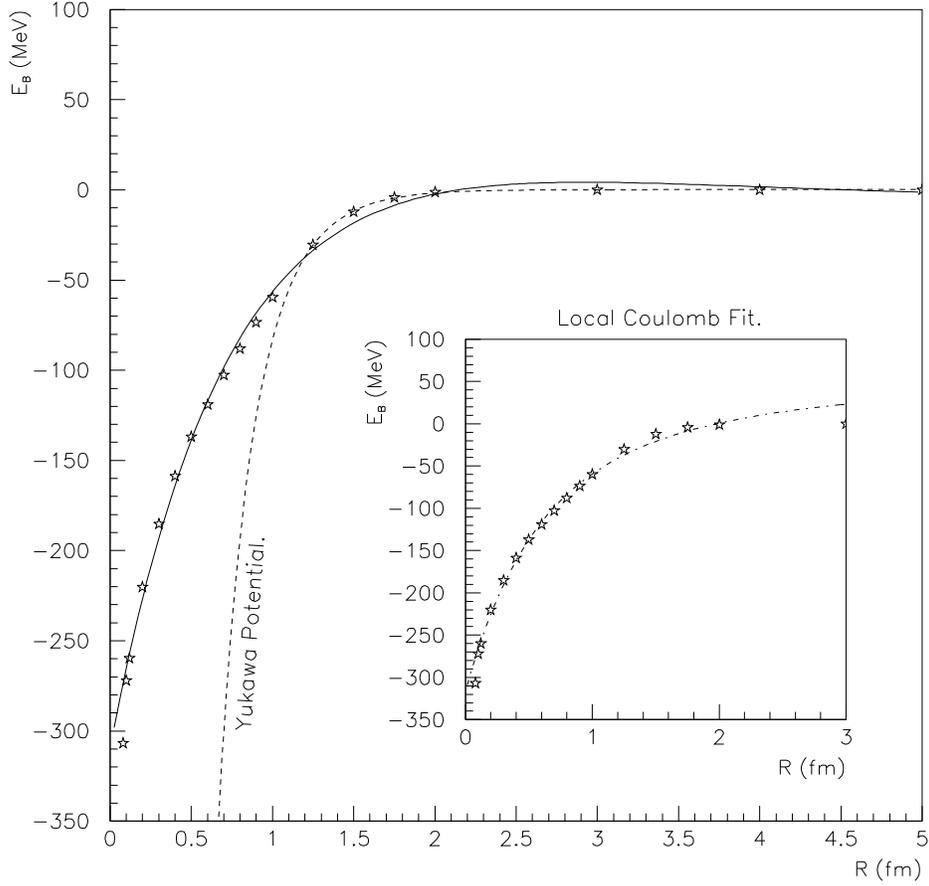}}
\end{center}
\caption[foo]{\footnotesize The   $Q_2$ binding energy curve   using the
         full flux-bubble interaction, {\it  i.e.}, in which the  colour
         is allowed to move around.   This curve was parameterized using
         $u(r)$ given in Fig.~\ref{fig:oldwf}.}
\label{fig:modelde}
\end{figure}

 The Monte Carlo computations  that were done, using the pseudo-hydrogen
wave function,  $\tilde\Psi_H$  ({\it  i.e.},   Eq.~\ref{eq:watwav})  in
\S~\ref{sec-old},      have     been    repeated    here     for     the
pseudo-hydrogen-molecular wave function, $\tilde\Psi_{H_2}$ ({\it i.e.},
Eq.~\ref{eq:psdo}),   and are    shown in    Figs.~\ref{fig:newwf}   and
\ref{fig:modelde}.

  It can be immediately seen, that there  is a dramatic contrast between
the figures  for  $\tilde\Psi_H$ and $\tilde\Psi_{H_2}\,$. The  depth of
the pseudo-Morse potential has increased from around 5  MeV to around 80
MeV. This  is sufficient  to  bind quark molecules.  From  the plots  in
Fig.~\ref{fig:newwf} we find that heavy quarks with
\begin{equation}
\mu_Q\approxge(660\pm24)\,MeV\,.
\end{equation}
such as $c$ and $b$ would form bound quark molecules.
We can fit  a Yukawa potential to the tail of the potential to give an 
effective Yukawa mass
\begin{equation}
\bar m_{\rm ex}\approx (575\pm32)\,MeV\,,
\end{equation}
{\it via} table~\ref{tb:yuk}, which is  still high compared to the  pion
mass.  Also   included  in Fig~\ref{fig:newwf},  are  plots for $\alpha$
fixed at  $2$  and $1.74$.  The subtle effects  seen  with the  old wave
function are simply overwhelmed by the  depth of the potential: in fact,
all  of the curves  are independent  of the   details of  the potential.
Therefore, any  new    model  of  nuclear   matter   that   incorporates
$\tilde\Psi_{H_2}$ should run into no difficulties by fixing $\alpha$.

 The final figure, Fig.~\ref{fig:modelde}, of the flux-bubble model with
moving   colour    is  quite        intriguing.   The  anomalies      in
Fig.~\ref{fig:modelda}  have  disappeared; the   light quarks   have not
drifted away as an isolated pair to leave a linear potential between the
heavy quarks. Na\"{\i}vely,  this seems to  suggest that the problem was
with the wave function and not $SU_c(2)$. However, this is not quite the
case, since the old wave function gives an  interior well depth which is
twice as deep. Therefore, it is still  energetically more preferable for
the $Q_2$ system to  dissociate into two  isolated mesons; one  with two
light quarks and  the other with two  heavy  ones.  The interior  of the
well can be fitted to a Coulomb potential of the form,
\begin{equation}
V_C(r)=-\frac{a}{r}\,\left(1-\dspexp{-\beta r}\right)+V_\infty
\end{equation}
(see         insert        in        Fig.~\ref{fig:modelde}),       with
$a\approx(135.5\pm2.9)\,MeV\,fm$,
$\beta\approx(2.844\pm0.048)\linebreak[0]fm^{-1}$,
$V_\infty\approx(68.3\pm2.6)\,MeV$. The term in the brackets is included
to mimic the  overlap between  the charge  distributions of two  mesonic
systems. In terms of $\alpha_s$,  the hyperfine constant for this region
of the potential is
\begin{equation}
\alpha_{Q_2}\approx(6.86\pm0.15)\alpha_s\,.
\end{equation}
The interior part  of  the potential is quite  deep  and bottoms out  at
${\cal O}(-270)\linebreak[0]MeV$, at  which point the $-\alpha_s/r$ term
for  the  heavy quarks  kicks in  ({\it  i.e.},  for $R\le0.1\,fm$). The
exterior  part    of the  potential  fits to    a  Yukawa potential with
$m_{\mbox{\tiny     ex}}\approx(612\pm32)\,MeV$         ({\it       via}
table~\ref{tb:yuk}).

\section{Discussion}
\label{sec-chpdsc}

  Various aspects  of model    building for  nuclear  matter  have  been
examined in the context of the $Q_2$  system. The ramifications of these
investigations will now be discussed.  The most  important result is the
new variational wave function, Eq.~\ref{eq:psdo}. This new wave function
makes a  massive  change in   the  depth of the   $Q_2$ potential  well,
increasing it by    a factor of $27$,   which  is deep   enough  to bind
relatively light quarks:  {\it i.e.},  $m_q\approxge{\cal O}(m_s)$.  The
wave function also    fulfills    the requirement  of handling     local
flux-bubble  interactions,  which becomes  apparent  when looking at the
``before''   and    ``after''    pictures    (of  moving    colour)   in
Figs.~\ref{fig:modelda} and \ref{fig:modelde}, respectively.  Therefore,
in $SU_\ell(2)$ for a many quark system this would suggest the following
$ansatz$:                    
\begin{equation}               
\Psi\sim{\rm     Perm}|\tilde\Psi_H(r_{p_ip_j})|\,    
\prod_{\mbox{\tiny colour}}|\Phi(r_{p_k})|
\label{eq:ansatz}
\end{equation}
where
\begin{equation}
\tilde\Psi_H(r_{p_ip_j})=\dspexp{-(\beta r_{p_ip_j})^\alpha}\,,
\end{equation}
${\rm    Perm}|\tilde\Psi_H(r_{p_ip_j})|$     is   a  totally  symmetric
pseudo-hydrogen wave function     and $|\Phi(r_{p_k})|$  is  a   totally
antisymmetric Slater wave function.

  For the full three-quark system  a similar wave function would  apply.
This does not necessarily  mean that this wave   function would lead  to
nuclear binding, but,  given the order of magnitude  of increase in  the
$Q_2$ well depth it would seem quite plausible that it might be a strong
enough effect  to produce a shallow   well in the nuclear-binding-energy
curve.   A simple  test  would be to  consider  $SU_\ell(2)$ with just a
string-flip  potential,  with $\alpha$ fixed,   in which case scaling is
restored and the Monte Carlo becomes quite straightforward to do.

  When  the  $SU_\ell(2)$  flux-bubble  model  with  moving  colour  was
considered, the  results showed that the  $Q_2$ system  dissociated into
one light  and one  heavy meson.  However, this  model is not  physical:
rather we should  consider the heavy quarks  as a composite of two light
quarks and use   $SU_c(3)$ instead,  so that  a   flux-tube cannot  form
between  the  two heavy   quarks.   For this   ``composite'' model,  the
$\lambda_{p_ip_j}$'s of potential Eq.~\ref{eq:stringyb} become
\begin{equation}
\lambda_{p_ip_j}=\left\{\begin{array}{rl}
 \frac{1}{3}& {\rm if}\;p_ip_j\,
       \varepsilon\,\{bb,\bar B\bar B\}\\
 \frac{1}{6}& {\rm if}\;p_ip_j\,
       \varepsilon\,\{b\bar G,\bar Bg\}\\
-\frac{1}{6}& {\rm if}\;p_ip_j\,
       \varepsilon\,\{b\bar B,g\bar G\}\\
-\frac{1}{3}& {\rm if}\;p_ip_j\,
       \varepsilon\,\{bg,\bar B\bar G\}\\
-\frac{4}{3}& {\rm if}\;p_ip_j\,
       \varepsilon\,\min\{q_r\bar q_s\}
\end{array}\right.\,,
\end{equation}
such  that $rg\sim\bar B$  and $rb\sim\bar G$  for the composite states,
$bb\sim gg$ and $\bar B\bar B\sim\bar G\bar  G$. If the heavy quarks are
considered to be a composite of two  light $u$ quarks then the effective
mass would be  about 600  MeV, which is   a little too small to  produce
binding    in this  model,  but    for masses  this  low the   adiabatic
approximation would  no   longer be   valid.  Therefore, an  interesting
possibility would  be    to   consider  a many-body    $SU_\ell(2)$   or
$SU_\ell(3)$  flux-bubble model in which  there  is an imbalance between
the  quark  and anti-quark masses.   An interesting side   effect of the
improved wave-function  is that it   will automatically produce  nucleon
deformations,  which,  it   has been    argued,   produce the    contact
interactions responsible for nuclear binding ~\cite{kn:Siemens}.

 Further enhancements are  expected by going to  the full $qqq$  nucleon
model, followed  by including flux-bubble interactions.   In particular,
the colour-hyperfine   interaction   which is  shown  by linked  cluster
expansion   models to  play   a significant  role   in nuclear processes
~\cite{kn:Nzar}.   The  flux-bubble   model  proved quite successful  at
combining colour-Coulomb        interactions         with      flux-tube
interactions. Although these interactions  had very little effect on the
$Q_2$   system,  it  was  useful   in  demonstrating that extending  the
flux-tube model to include  local  perturbative QCD interactions  can be
done.   This  is important, since  the    nuclear hard-core potential is
usually  believed    to have  its    origin   in hyperfine  interactions
\cite{kn:Boyce,kn:thesisb,kn:Nzar}.  Therefore,  it   would prove   most
interesting to investigate     the effect of  adding  more  perturbative
interactions to  the $Q_2$ system.   It  also appears that  relativistic
effects are unimportant.  With the addition of $SU_c(3)$ this would lead
to a more  realistic model of mesonic  molecules which  perhaps could be
tested in the laboratory.

  The $Q_2$ system has proven to be a very useful aid for trying to sort
out  the complexities of model  building for nuclear matter. The details
of the mechanics, from wave functions to dynamics to practical computing
methods,    of the   flux-bubble   model    have now been     thoroughly
investigated. It appears that the flux-bubble model may prove to be very
successful,  not only for modeling  nuclear matter but also for modeling
mesonic molecules as well.

\noindent{\bf Acknowledgments}

 This  work was  supported by NSERC.  The  Monte Carlo calculations were
performed  on an 8  node DEC 5240  UNIX CPU farm, using Berkeley Sockets
(TCP/IP) \cite{kn:Stevens}.   The computing  facilities were provided by
the   Carleton University   Department  of   Physics,  OPAL,   CRPP, and
T{\scriptsize  HEORY} groups.  We   would like to thank  M.A.~Doncheski,
H.~Blundell,     M.~Jones, and  J.S.~Wright   for   useful  discussions.
\appendix
\section{Analysis}
\label{sec-anal}

 This appendix contains a summary of the analysis done for the models
given in sections \ref{sec-old} through \ref{sec-new} of this paper.

  All of the results of these sections make use of the general potential
Eq.~\ref{eq:fixed}.    For        $\bar  U(R)$,     as      defined   by
Eq.~\ref{eq:effective}, some  analytical  results were  obtainable where
the    $Q_2$ system effectively   dissociates  into  two  isolated meson
systems. For all of the plots  this occurs at  $\bar U(\infty)$, and for
the  linear  and linear-plus-Coulomb  plots  this  also occurs  at $\bar
U(0)$. In  these regions,  the  analytic  expression is  given  by ({\it
cf}. \cite{kn:Watson})
\begin{eqnarray}
E_{free}&=&2
\left[
 \frac{g_T^{\mbox{}}(\alpha)}{2\mu}\,\beta^2+
 \sigma
 \left(
  \frac{g_L^{\mbox{}}(\alpha,\beta)}{\beta}-
  g_0^{\mbox{}}(\alpha,\beta)\,r_0
 \right)\right.\nonumber\\\nonumber\\&&\left.-
 \frac{3}{4}\,\alpha_s
 \left(
  g_C^{\mbox{}}(\alpha,\beta)\,\beta-
  \frac{(1-g_0^{\mbox{}}(\alpha,\beta))}{r_0}
 \right)
\right]\,,\label{eq:anab}
\end{eqnarray}
where
\begin{eqnarray}
g_L^{\mbox{}}(\alpha,\beta)&=&
[1-{\rm P}(4/\alpha,2(\beta r_0)^\alpha)\,]\,g_L^{\mbox{}}(\alpha)\,,
\;\;\;\mbox{}\\
g_C^{\mbox{}}(\alpha,\beta)&=&
{\rm P}(2/\alpha,2(\beta r_0)^\alpha)\,g_C^{\mbox{}}(\alpha)\,,\\
g_0^{\mbox{}}(\alpha,\beta)&=&
1-{\rm P}(3/\alpha,2(\beta r_0)^\alpha)\,,
\end{eqnarray}
such that
\begin{eqnarray}
g_T^{\mbox{}}(\alpha)&=&
\frac{\alpha^2\,\mbox{2\raisebox{1.5ex}{$\frac{2}{\alpha}-2$}}
\,\Gamma(2+1/\alpha)}{\Gamma(3/\alpha)}\label{eq:anaaa}\,,\\
g_L^{\mbox{}}(\alpha)&=&
\frac{\Gamma(4/\alpha)}{
\mbox{2\raisebox{1.5ex}{$\frac{1}{\alpha}$}}\,\Gamma(3/\alpha)}
\label{eq:anaab}\,,\\
g_C^{\mbox{}}(\alpha)&=&
\frac{\Gamma(2/\alpha)}{
\mbox{2\raisebox{1.5ex}{$-\frac{1}{\alpha}$}}\,\Gamma(3/\alpha)}\,,
\end{eqnarray}
${\rm P}(a,z)=1-\Gamma(a,z)/\Gamma(a)\,$,    and   $\Gamma(a,z)$ is  the
incomplete   gamma  function  \cite{kn:Wolfram,kn:Abramowitz}.   In  the
limits         as    $r_0\rightarrow0$    and     $r_0\rightarrow\infty$
Eq.~(\ref{eq:anab}) reduces to solutions for the purely linear,
\begin{equation}
E_{free}\;\;
\mbox{\raisebox{-0.6ex}{$
 \stackrel{\longrightarrow}{
 \mbox{\tiny$r_0\rightarrow0$}}$}
}
\;\;2\left(\frac{g_T^{\mbox{}}(\alpha)}{2\mu}\,\beta^2+
g_L^{\mbox{}}(\alpha)\,\frac{\sigma}{\beta}\right)\,,
\;\;\;\;\;\;\;
\label{eq:anaa}
\end{equation}
and purely Coulomb,
\begin{equation}
E_{free}\;\;
\mbox{\raisebox{-0.6ex}{$
 \stackrel{\longrightarrow}{
 \mbox{\tiny$r_0\rightarrow\infty$}}$}
}
\;\;2\left(\frac{g_T^{\mbox{}}(\alpha)}{2\mu}\,\beta^2-
\frac{3}{4}\,\alpha_s g_C^{\mbox{}}(\alpha)\,\beta\right)\,,
\end{equation}
cases, respectively.  Table~\ref{tb:bart}  gives a summary of  the Monte
Carlo vs. analytic results.
\begin{table}[htbp]
\caption[foo]{ \footnotesize Monte Carlo (MC) vs. analytic results for
  Figs.~\ref{fig:oldwf}, \ref{fig:modelda}, \ref{fig:newwf} and
  \ref{fig:modelde}.}
\begin{center}
\begin{tabular}{@{}llccccc@{}}\hline
Method & & Fig. & $R\;(fm)$ & $E_{free}\;(MeV)$
& $\alpha$ & $\beta\;(fm^{-1})$\\\hline\hline
Analytic & Eq.~\ref{eq:anaa} & n.a.
& $0/\infty$ & 1709.61   & 1.75 & 1.37 \\\hline
MC & Linear
 & \ref{fig:oldwf} & 0 & $1709.424\pm0.038$ & 1.74 & 1.37 \\
& &     & 5 & $1709.492\pm0.044$ & 1.72 & 1.38 \\\cline{3-7}
& & \ref{fig:newwf} & 0 & $1709.647\pm0.032$ & 1.77 & 1.36 \\
& &     & 5 & $1709.485\pm0.032$ & 1.75 & 1.36 \\\hline\hline
Analytic & Eq.~\ref{eq:anab} & n.a.
& $0/\infty$ &  1527.07  & 1.74 & 1.37 \\\hline
MC & $\;\;$Linear
 & \ref{fig:oldwf} & 0 & $1527.171\pm0.029$ & 1.78 & 1.36 \\
& $\;\;\;\;\;$+
 &     & 5 & $1527.171\pm0.029$ & 1.78 & 1.36 \\\cline{3-7}
& Coulomb
 & \ref{fig:newwf} & 0 & $1527.884\pm0.041$ & 1.74 & 1.34 \\
& &     & 5 & $1527.884\pm0.041$ & 1.74 & 1.34 \\\hline
MC & Flux-Bubble$\;\;$
 & \ref{fig:oldwf} & 5 & $1526.974\pm0.032$ & 1.74 & 1.36 \\\cline{3-7}
& & \ref{fig:modelda} & 5 & $1527.884\pm0.041$ & 1.74 & 1.34
\\\cline{3-7}
& & \ref{fig:newwf} & 5 & $1527.884\pm0.041$ & 1.74 & 1.34 \\\cline{3-7}
& & \ref{fig:modelde} & 5 & $1527.173\pm0.062$ & 1.68 & 1.40 \\\hline
\end{tabular}
\end{center}
\label{tb:bart}
\end{table}
The   results for  the  linear   and linear-plus-Coulomb potentials,  at
$R=0\,fm$ and $R=5\,fm$   ({\it  i.e.}, $\approx\infty$),  were  checked
against  Eqs.~\ref{eq:anaa}   and  \ref{eq:anab},  respectively.     The
flux-bubble  case, at $R=5\,fm$,  was verified  using Eq.~\ref{eq:anab}.
The minima  of the analytic  expressions, which were  used to verify the
aforementioned    models,   were     found   {\it   via}      the   {\tt
FindMinimum[\ldots]} routine in Mathematica \cite{kn:Wolfram}.

  The binding energy  of the $Q_2$ system  can be  estimated by doing  a
local  parabolic fit \cite{kn:Flugge}  about the minimum of $\bar U(R)$:
{\it i.e.}, by fitting
\begin{equation}
y(r)={\cal C}(r-r_0)^2-{\cal D}
\label{eq:parabolic}
\end{equation}
such that ${\cal  C}=\mu_Q\omega^2/[2(\hbar c)^2]$, where $\mu_Q$ is the
reduced mass of   the heavy quarks.  Therefore,  the  binding  energy is
simply
\begin{equation}
E_h^\nu= (\nu+\mbox{$\textstyle\frac{1}{2}$})\hbar\omega-{\cal D}\,,
\end{equation}
which implies
\begin{equation}
\mu_Q\ge\mu_{\rm min}\equiv\frac{{\cal C}(\hbar c)^2}{2{\cal D}^2}
\label{eq:parbnd}
\end{equation}
in order to obtain binding. Table~\ref{tb:sue} shows the results for the
parabolic fits to the $\bar U(R)$'s in Fig.~\ref{fig:oldwf}.
\begin{table}[htbp]
\caption[foo]{\footnotesize Results for local parabolic fitting about
    the minima of the plots in Figs.~\ref{fig:oldwf} and
    \ref{fig:newwf}.}
\begin{center}
\begin{tabular}{@{}lccccc@{}}\hline
Model& Fig. & ${\cal C}\,(MeV/fm^2)$ & $r_0\,(fm)$&${\cal D}\,(MeV)$ &
$\mu_{\min}\,(GeV)$\\\hline
Linear  & \ref{fig:oldwf} &
 $29.8\pm1.5$ & $0.4174\pm0.0056$ & $3.268\pm0.058$ & $54.3\pm 3.3$\\
   & \ref{fig:newwf} &
 $234.8\pm8.0$ & $0.5288\pm0.0057$ & $82.55 \pm0.49$ & $671
\pm24$\\\hline
Linear + & \ref{fig:oldwf} &
 $28.7\pm1.4$ & $0.4134\pm0.0045$ & $3.098\pm0.060$ & $58.2\pm 3.6$\\
Coulomb & \ref{fig:newwf} &
 $230.3\pm8.0$ & $0.5833\pm0.0058$ & $82.60 \pm0.49$ & $657
\pm24$\\\hline
Flux-Bubble& \ref{fig:oldwf} &
 $20.0\pm1.0$ & $0.4614\pm0.0034$ & $2.783\pm0.042$ & $50.3\pm 2.9$\\
   & \ref{fig:newwf} &
 $226.8\pm8.0$ & $0.5840\pm0.0059$ & $82.24 \pm0.49$ & $653 \pm 24$
\\\hline
\end{tabular}
\end{center}
\label{tb:sue}
\end{table}

  The  asymptotic parts  of  the $\bar U(R)$'s  were  fitted to a Yukawa
potential of the form
\begin{equation}
V_Y(r)=-2f^2\frac{e^{-\mu r}}{r}\,.
\end{equation}
Table~\ref{tb:yuk} summarizes the Yukawa fits for all of the models.
\begin{table}[htbp]
\caption[foo]{\footnotesize Results for asymptotic Yukawa fits of the
    plots  in Figs.~\ref{fig:oldwf}, \ref{fig:newwf}, and
    \ref{fig:modelde}, where $m_{\rm ex}=\hbar c\mu$.}
\begin{center}
\begin{tabular}{@{}lcccc@{}}\hline
Model& Fig. & $f^2\,(MeV\,fm)$ & $\mu\,(fm^{-1})$ & $m_{\rm
ex}\,(MeV)$\\\hline
Linear & \ref{fig:oldwf} &
  $2.49\pm0.20$ & $3.23\pm0.21$ & $637\pm\;\;41$\\
  & \ref{fig:newwf} & $27.3\pm2.7$ & $2.97\pm0.16$ & $586\pm32$\\\hline
Linear+Coulomb &\ref{fig:oldwf} &
  $2.0\;\;\pm2.2\;\;$& $3.0\;\;\pm3.5\;\;$ & $590\pm690$\\
    & \ref{fig:newwf} &
  $25.9\pm2.7$ & $2.89\pm0.16$ & $570\pm32$\\\hline
Flux-Bubble & \ref{fig:oldwf}&
  $1.8\;\;\pm2.3\;\;$& $2.4\;\;\pm3.5\;\;$ & $470\pm690$\\
    & \ref{fig:newwf} &
  $25.9\pm2.7$ & $2.89\pm0.16$ & $570\pm32$\\
    &\ref{fig:modelde} &
  $30.3\pm2.1$ & $3.10\pm0.11$ & $612\pm32$\\
\hline
\end{tabular}
\end{center}
\label{tb:yuk}
\end{table}

\end{document}